\documentclass[,final]
  {aipproc}

\layoutstyle{8x11single}

\newcommand\doingARLO[2][]{%
  \ifx\mmref\undefined #1\else #2\fi
}
\newcommand{\nat}{Nature}
\newcommand{\apj}{ApJ}
\newcommand{\apjl}{ApJL}
\newcommand{\aj}{AJ}
\newcommand{\mnras}{MNRAS}
\newcommand{\aap}{A\&A}
\newcommand{\araa}{ARA\&A}
\newcommand{\physrep}{Physics Reports}
\begin{document}

\title 
      [GRB Afterglows]
      {Can optical afterglows be used to discriminate between Type I and Type II GRBs?}


\author{D. A. Kann}{
  address={Th\"uringer Landessternwarte Tautenburg,  Sternwarte 5, D--07778 Tautenburg, Germany},
%  email={kann@tls-tautenburg.de},
%  thanks={This work was commissioned by the AIP}
}

\copyrightyear{2008}

\begin{abstract}
The precise localization of short/hard (Type I) gamma-ray bursts (GRBs) in recent years has answered many questions but raised even more. I present some results of a systematic study of the optical afterglows of long/soft (Type II) and short/hard (Type I) GRBs, focusing on the optical luminosity as another puzzle piece in the classification of GRBs.
\end{abstract}

\date{\today}

\maketitle

\section{Introduction}

In the last decade, since the precise localization via the discovery of their optical \citep{vanParadijs970228}, X-ray \citep{Costa970228} and radio \citep{Frail970508} afterglows, much knowledge has been gained about the high-energy transients known as gamma-ray bursts, which we now know are extremely energetic \citep{Kulkarni990123} explosions of massive stars \citep{WoosleyBloom} in cosmological distances \citep{Metzger970508}. But there are two classes of GRBs \citep{Kouveliotou1993}, and those of short duration remained enigmatic until the advent of the dedicated GRB mission \emph{Swift} \citep{GehrelsSwift}. This satellite made it possible to also pinpoint short GRBs via their afterglows in the X-ray \citep{Gehrels050509B}, optical \citep{Hjorth050709, Fox050709} and radio \citep{Berger050724} regimes, establishing not only that they too originate at cosmological distances, but also that at least some of them are not associated with recent star formation \citep{Berger050724, Barthelmy050724}, implying that they must originate from different progenitors than long/soft GRBs. The most favored model is the coalescence of two compact objects of which at least one is a neutron star \citep[for reviews, see][]{NakarReview, LeeRR}.

The discovery of two temporally long GRBs with no associated supernova emission down to deep limits \citep{Fynbo060614, Gal-Yam060614, DellaValle060614, Ofek060505} exacerbated the debate on GRB classification \citep{Gehrels060614} which had been triggered by GRB 050724, a GRB in an elliptical galaxy which was temporally long as seen by \emph{Swift}, but would have been classified as short if detected by BATSE, due to the presence of a long, soft, X-ray-bright tail of emission. This led \cite{Zhang060614}, in the light of classifying GRBs according to the underlying progenitor physics \citep{BloomPhysics}, to create a new classification based on several observed quantities and phenomena, such as duration, spectral lag \citep{NorrisBonnell}, light curve shape and host galaxy characteristics. GRBs that originate in the explosions of massive stars are called Type II events, whereas those unassociated with recent star formation (probably due to compact object coalescence) are Type I GRBs.

By the end of 2007, the \emph{Swift} mission had resulted in a plethora of (optical) afterglow discoveries, allowing the creation of a large sample of Type II GRB afterglows, as well as a first moderately large sample of Type I GRB afterglows. Since several years, I and my collaborators have been systematically studying large samples of optical afterglows. In this article, I report results from two recent papers, the first one \citep{KannPaperIV} comparing the afterglows of \emph{Swift}-era Type II GRBs with those of the pre-\emph{Swift} era \citep{KannPaperIII}, the second one comparing the afterglows of Type I GRBs to the complete Type II GRB afterglow sample \citep{KannPaperV}. I focus especially on three enigmatic events, GRBs 060121, 060505 and 060614, which point to problems in the Type I/II classification scheme, in the light of the luminosities of their optical afterglows.

\section{Afterglows of Type I and Type II GRBs}

\begin{figure}
\caption{The intrinsic afterglows of Type II GRBs. All have been corrected for extinction (both Galactic and rest-frame) and shifted to a common redshift of $z=1$ to allow a direct comparison. The thin gray lines mark the pre-\emph{Swift} sample of \cite{KannPaperIII}, whereas the three other colors denote afterglows from the sample of \cite{KannPaperIV}. Red: Golden Sample; Blue: Silver Sample; Black: Bronze Sample. The vertical lines mark three points in time where the optical luminosity was examined, at 43 seconds, 12 hours and 2 days in the rest-frame. The dash-dotted line with decay slope $\alpha=1.2$ marks, at late times, the maximum luminosity trend. Bursts of note are named.}
\includegraphics[height=1\textwidth]{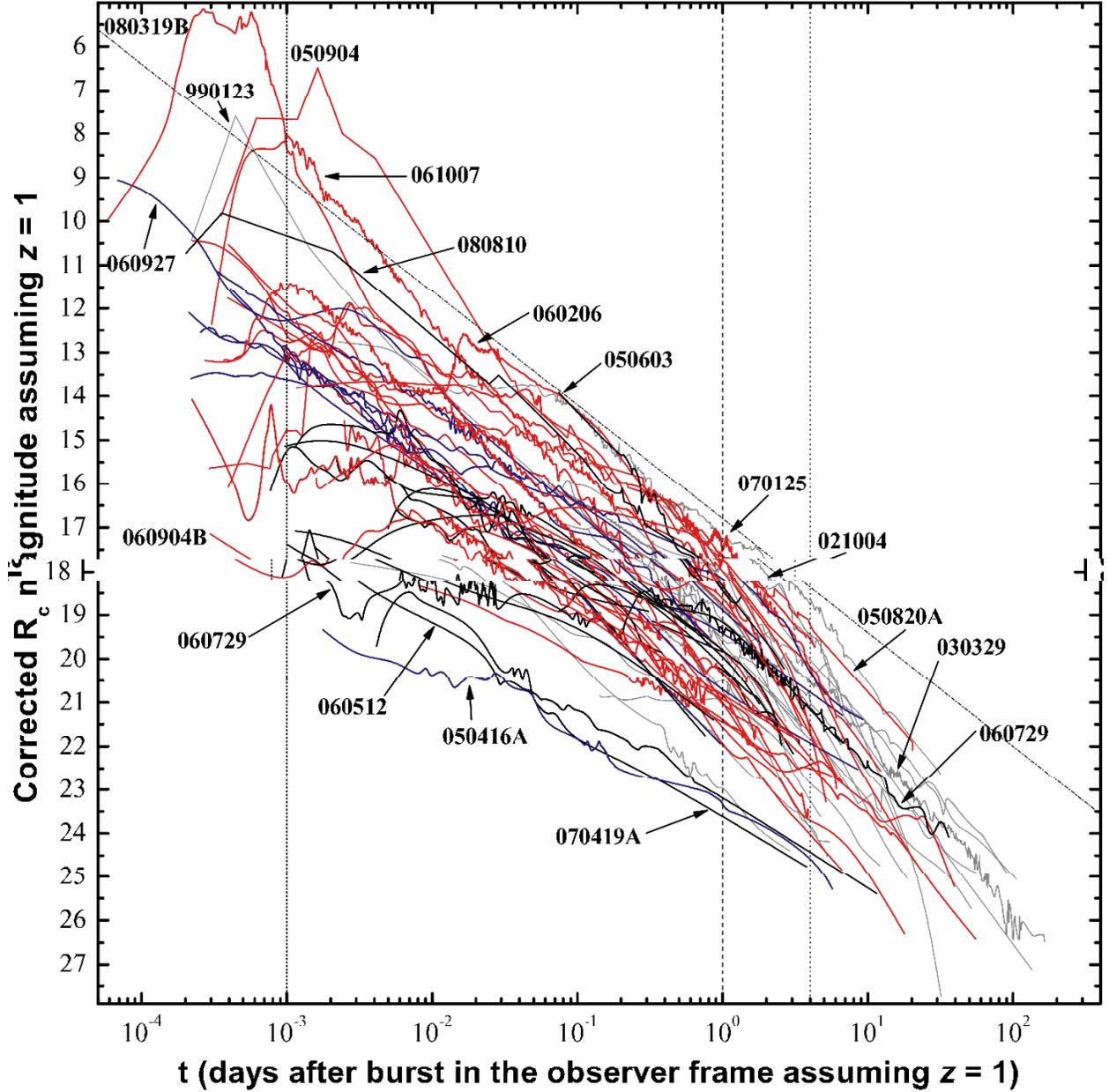}
\end{figure}

Kann et al. \citep{KannPaperIII} presented a systematic analysis of pre-\emph{Swift} GRB afterglows (Type II GRBs only) based on a maximized photometry sample. For a ``Golden Sample'' of 19 GRBs, spectral energy distributions (SEDs) could be constructed which allowed a fit that left both the intrinsic spectral slope $\beta$ ($F_\nu\propto\nu^{-\beta}$, GRB afterglow emission in the optical/NIR regime is almost always a simple power-law spectrum) as well as the rest-frame extinction $A_V$ as free parameters. The dust absorption was modeled with the dust extinction laws of the Milky Way (MW), Large (LMC) and Small Magellanic Clouds (SMC).

Next to deriving the statistics on the spectral slope and the extinction, the main result of this work was the derivation of the optical luminosities of GRB afterglows at fixed rest-frame times. Using the spectral slope and the rest-frame extinction, I was able to shift all afterglows to a common redshift of $z=1$, fully corrected for foreground absorption. I found that at 0.5 rest-frame days, the optical luminosities clustered, their distribution less wide than the observed one, and that the afterglows of nearby GRBs were fainter than those of more distant GRBs \citep[see also][]{LiangZhang, Nardini2006, Nardini2008A}.

\subsection{Comparing pre-\emph{Swift} and \emph{Swift}-era Type II GRB afterglows}

The work on Type II GRB afterglows was extended into the \emph{Swift} era by \cite{KannPaperIV}. The sample discussed in this work comprises almost 50 events, including some detected by other satellites (\emph{INTEGRAL}, \emph{HETE II} or the \emph{IPN}). To make the sample as inclusive as possible, I relaxed the ``Golden Sample'' criteria and created two further samples, Silver and Bronze, with the Bronze Sample afterglows not corrected for intrinsic extinction, as the SED did not allow this value to be derived \citep[for more details on the sample selection criteria, see][]{KannPaperIV}.

The main result of this study is that there are no fundamental differences between the afterglow samples from the pre-\emph{Swift} and the \emph{Swift} era (see Fig. 1). The perceived dimness of \emph{Swift} GRB afterglows \citep{BergerSwiftAfterglows} is mostly due to the higher redshifts \cite{JakobssonSwiftRedshifts} of these GRBs. Further results are:

\begin{itemize}
\item The prevalence for dust similar to that of the SMC is confirmed
\item A low-significance trend of lower extinction at higher redshifts is strongly confirmed by the addition of many GRBs at $z\ge3$
\item The typical extinctions $A_V$ along the line of sight are still $\approx0-0.2$ magnitudes
\item The luminosity distribution of the Bronze Sample GRBs becomes identical to those of the Golden and Silver samples (which are very similar) if a typical extinction is assumed to affect these afterglows
\item The clustering of luminosities as well as the two afterglow populations (faint, near vs. bright, distant) are confirmed \citep[see also][]{Nardini2008B}
\item A possible exponential cutoff in the luminosity function for the brightest afterglows is found
\item More than 50\% of all afterglows already detected at 43 seconds after the beginning of the GRB in the rest frame cluster within two magnitudes
\item Only a few exceptional events are significantly brighter at early times, reaching up to naked-eye visibility even if at $z=1$ \citep[see also][]{Kann050904, Bloom080319B, Racusin080319B}
\item A low-significance correlation between the bolometric isotropic energy release in the prompt emission and the afterglow luminosity at late times is seen
\item Recent redshift results from host galaxy studies reveal a new population of optically dark low-luminosity GRBs at low redshifts
\end{itemize}

To conclude, the strong similarities between the Type II afterglows of all samples allows me to combine them as a single reference sample against which I can compare the Type I GRB afterglows.

\subsection{Comparing Type II and Type I GRB afterglows}

\begin{figure}
\caption{The intrinsic afterglows of Type I GRBs (red) in comparison with those of Type II GRBs (gray). Upper limits are given by downward-pointing triangles connected with straight lines, detections are larger squares connected by splines. Clearly, the Type I GRB afterglows as a whole are fainter than most of the Type II GRB afterglow, I find a mean difference of five magnitudes between the two samples. I highlight the three GRBs discussed in more detail in the text. GRB 060121 (black) is seen to be of comparable luminosity to typical Type II GRB afterglows for the high redshift case (upper), and comparable to the fainter Type II GRB afterglows for the lower redshift case (lower).  Note the large deviations from a typical smooth power law decay for this afterglow. The afterglow of GRB 060614 (green), after an early detection, becomes fainter quickly, only to rise slowly in magnitude for a long time. It exhibits a clear break which can be interpreted as a jet break, and a smooth late decay without any sign of a supernova. GRB 060505 (blue) is only detected at a single epoch (at about 2 days after the GRB in the $z=1$ frame), may rise initially, and must exhibit a steep late-time decay. The final limit is the most stringent one found so far on a supernova contribution.}
\includegraphics[height=1\textwidth]{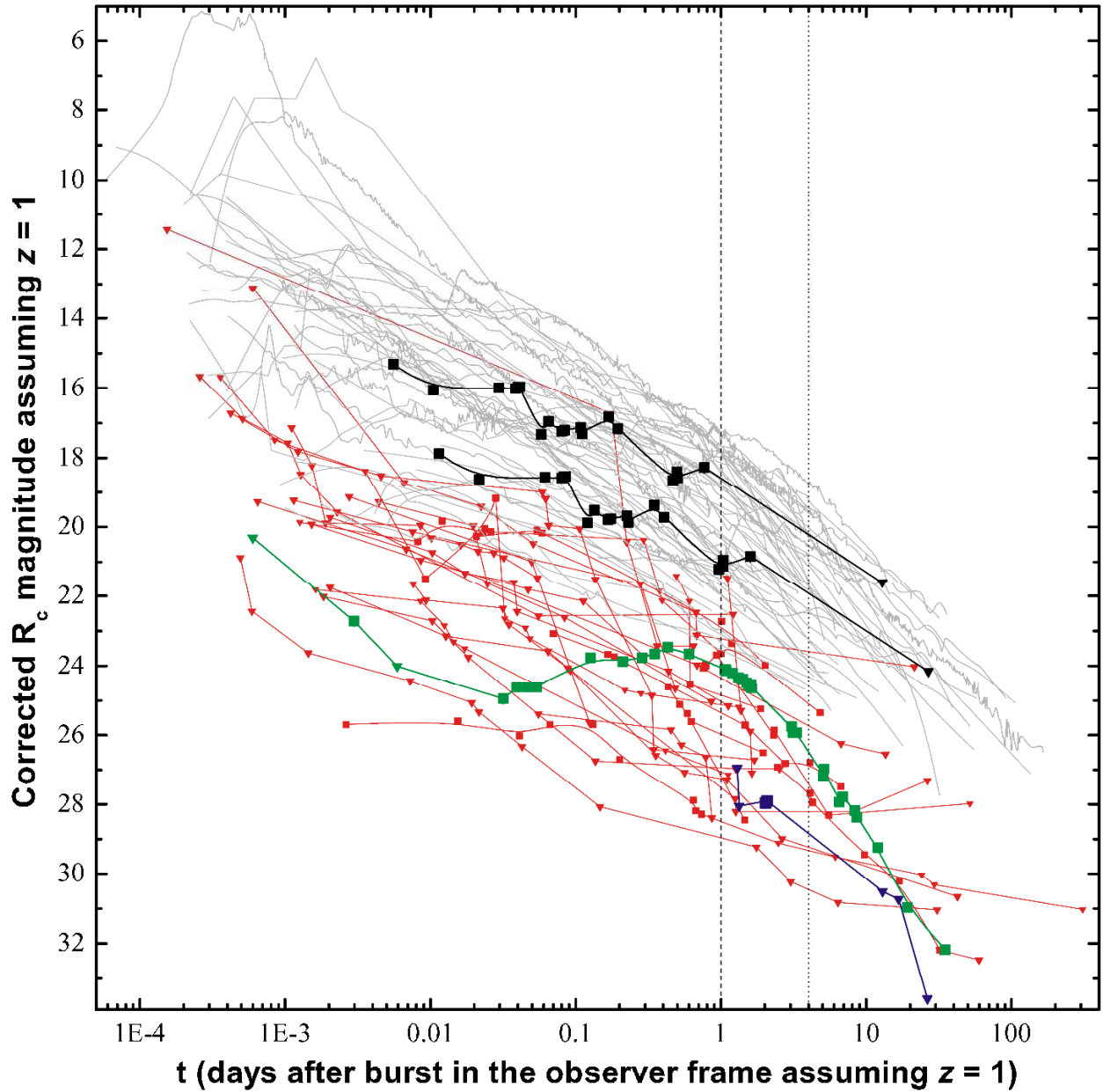}
\end{figure}

The detailed analysis of the Type I GRB sample has been presented in \cite{KannPaperV}. See there especially for the details of the sample selection. For these afterglows, I incorporate not only detections but also upper limits, and also afterglows with uncertain (or simply none) redshifts. This allows me to maximize the sample size, but I must be cautious in deriving results. My main result is that the afterglows of Type I GRBs are significantly fainter than those of Type II GRBs; in the mean, they are five magnitudes fainter, that is, a factor of 100 in flux density. The addition of upper limits implies that the mean magnitude difference may be even larger. Further results from this work are:

\begin{itemize}
\item The intrinsic afterglows of Type I GRBs actually show an increased spread, not like the clustering seen in Type II GRB afterglows
\item No signatures of classical radioactivity-driven SNe are seen in those Type I GRB afterglows with late-time data, in a few cases down to extremely deep limits
\item The upper limits on any optical emission from GRB 050509B allow stringent constraints to be placed on the parameter space of an accompanying ``Mini-SN''
\item The addition of Type I GRB afterglows strengthens the correlation between bolometric isotropic energy release and late-time afterglow luminosity, but there is an offset between the two classes, pointing to the influence of the density of the circumburst medium
\item The brightest Type I GRBs and the faintest Type II GRBs in the samples overlap in the aforementioned diagram
\item For GRBs with secure redshifts, there is a clear trend visible between afterglow luminosity and host galaxy offset, once more interpreted as a lower circumburst density
\item There is no correlation between afterglow luminosity and the duration of the events, or between afterglow luminosity and the existence of an extended soft emission component
\item A plot of the optical luminosity versus redshift for all GRBs shows the sign of a detector-sensitivity cutoff
\end{itemize}

\section{GRBs 060121, 060505 and 060614 in the light of their optical afterglow luminosities}

I will now focus on three GRBs which do not fit easily into the Type I/II classification scheme, and add what I have learned about the afterglow properties of the two samples to see if I am able to strengthen the arguments for one or the other side of the debate.

\subsection{GRB 060121}

GRB 060121 was a bright, temporally short GRB ($T_{90}\le2$ seconds at high energies) localized by \emph{HETE II} \citep{Donaghy060121}. The prompt emission exhibited several features which have been found to be indicators of Type I GRBs \citep{NorrisBonnell}. Next to the short duration at high energies (in the BATSE energy range which was used to determine the original $T_{90}$ definition), the soft X-ray detectors on \emph{HETE II} detected a faint, soft and extended emission component such as had first been seen in GRB 050724 and multiple further events hence. Furthermore, the spectral lag of the initial prompt emission spike was negligible. Optical observations discovered \citep{KannPaperV} a very faint, red afterglow, and an extremely faint host galaxy (from which the afterglow was significantly offset), in whose neighborhood several Extremely Red Objects were found \citep{Levan060121, BergerHighzSGRBs}. Analysis of the spectral energy distribution indicated that this event lay either at very high redshift ($\approx4.6$) (the favored solution) or at lower redshift ($\approx1.7$), but with large intrinsic dust absorption \citep{deUgarte060121}; in any case, at a distance beyond even the newly found ``high redshift'' Type I GRBs, which is also supported by the host galaxy colors \citep{BergerHighzSGRBs}.

As it is, GRB 060121 was already one of the most energetic ``short GRBs'' observed; a redshift of four or higher would imply an isotropic energy release several orders of magnitude higher than typical Type I GRBs. Broadband analysis of the light curve has revealed evidence for a jet break \citep{deUgarte060121}, which would ease the required energy budget considerably; nonetheless, this would still remain an extraordinary event, especially since it calls for a rapid progenitor channel. Therefore, doubt has been cast on the derivation of the photometric redshift, or it is thought that this is a temporally very short Type II event, associated with the death of a massive star.

Upon correcting for extinction and distance, the very faint afterglow of GRB 060121 becomes much brighter (see Fig. 2). Indeed, for the $z=4.6$ solution, it is comparable to average Type II GRB afterglows. The $z=1.7$ solution (which corrects for much larger extinction) is still comparable to faint Type II GRB afterglows. Clearly, the optical afterglow luminosity is extraordinary in comparison to the mean Type I GRB afterglow luminosity, just as the isotropic energy release is an outlier. In the light of the trend seen between optical luminosity and prompt energy release, this result does not come as a surprise. Therefore, the optical afterglow does not really further the understanding of this event, but points to two possible extremes: If this is a Type I GRB, it is part of an ``high-energy tail'' orders of magnitude more energetic than typical events, possibly due to extreme collimation. If it is a Type II event, there is no problem with the energy budget, but an explanation must be found for the shape of the prompt emission, which would be extremely short ($\approx0.3$ seconds) in the rest frame, well in the region of classical ``short/hard'' GRBs.

\subsection{GRB 060505}

GRB 060605 was a moderately bright event localized via ground processing by \emph{Swift}, a flight-localization had been prevented by a high background due to SAA entry. Follow-up observations a day after the event revealed a faint optical and X-ray afterglow associated with a nearby ($z=0.0889$), cataloged spiral galaxy. Late-time observations detected neither an afterglow nor an expected supernova down to very stringent limits \citep{Fynbo060614, Ofek060505}. The event was suggested to be a Type I GRB, despite the duration of $\approx4$ seconds \citep{Ofek060505}, or a Type II GRB, despite the completely missing supernova \citep{Fynbo060614}. A detailed spectroscopic observation of the host galaxy \citep{Thoene060505} showed that the GRB had occurred in a massive star-forming region in one of the spiral arms, which, taken by itself, exhibited the typical characteristics (e.g. low metallicity, high specific star formation rate, blue color) of a Type II GRB host galaxy, indicating a massive star as a progenitor. Furthermore, the study of the prompt emission, as detected by \emph{Swift} and \emph{Suzaku HXD-WAM}, revealed not only a faint precursor (extending the duration to 10 seconds), but also a non-negligible spectral lag, which has not been found for unambiguous Type I GRBs \citep{McBreen060505}.

While only detected in a single epoch, high signal-to-noise multicolor observations of the afterglow were obtained which allow dust extinction to be ruled out as the cause for the faint supernova \citep[][this is also an argument against a background GRB]{Thoene060505}. Shifted to $z=1$, I find that the afterglow is exceedingly faint (see Fig. 2) and must also decay rapidly. From the perspective of optical luminosity, it seems more likely that this is a Type I GRB, but both the environment as well as the spectral lag speak against this interpretation (the duration is only a moderate counterargument). On the other hand, it has been shown that very fast channels exist to produce merging neutron stars within the lifetime of an HII region \citep{Belczynski}, and as the physical origin of spectral lag is yet unclear, there is no strong argument why a Type I GRB with significant spectral lag should not exist \citep{KannPaperV}. If this is a Type II event, then a novel explosion process such as a delayed-fallback black hole (in which the black hole is not created directly during core collapse, but through delayed accretion in a weak supernova explosion) must be invoked \citep{Fynbo060614, Fryer2006, Fryer2007}, which would have to be able to produce a relativistic jet (creating the GRB and the later afterglow) yet release only a very small amount of energy into the environment (faint prompt emission, faint afterglow, negligible production of radioactive elements to prevent the supernova).

\subsection{GRB 060614}

GRB 060614 was a very bright temporally long ($T_{90}=102$ seconds) GRB localized by \emph{Swift} and detected by many other satellites. Its afterglow was followed up extensively in the optical, and associated (at a moderate offset) with a very faint and moderately star-forming host galaxy at $z=0.125$. The very low redshift motivated an extensive search for an accompanying supernova, and similar to GRB 060505 a month earlier, none was found \citep{Fynbo060614, Gal-Yam060614, DellaValle060614}. The \emph{Swift} data allowed a detailed study of the prompt emission properties. While at first glance looking like a classical, spiky high luminosity Type II GRB, spectral analysis revealed that the GRB consisted of two parts, a hard initial spike complex of five seconds duration, and a much softer, though in this case still very energetic tail which produced the long duration. Both components show negligible spectral lag \citep{Gehrels060614}. The association with the host galaxy is secure, ruling out a background event \citep{Gal-Yam060614}, and while there is some dust found along the line of sight \citep{KannPaperV}, it can not be the explanation for the missing supernova.

In terms of both prompt energy release as well as optical afterglow luminosity, this event lies on the borderline between Type I and Type II GRBs (see Fig. 2), but it is not extraordinarily bright like GRB 060121. It was shown by \cite{Zhang060614} that the prompt emission, if reduced in fluence by about an order of magnitude, would strongly resemble that of GRB 050724, an unambiguous Type I event, and would have been seen as a ``short/hard'' GRB by BATSE. While there is no smoking gun, the moderate optical afterglow luminosity \citep[much smaller than that of the classical Type II GRB 030329, which occurred at a roughly similar distance,][]{KannPaperIII}, combined with the negligible spectral lag, prompt emission shape, missing supernova and host galaxy properties point to this event to be due to an energetic merger, and thus be of Type I.

\section{Conclusions}

I have presented the results of two studies on Type I and Type II GRB afterglows in the \emph{Swift} era, with a special focus on three enigmatic and hard-to-classify GRBs in the light of their optical afterglow luminosities. In none of the three cases is the afterglow luminosity a decisive factor in determining the GRB progenitor class, but they still offer a further puzzle piece toward a deeper understanding of GRBs and their progenitors.

\begin{theacknowledgments}
D.A.K. acknowledges B. Zhang for the invitation to the Nanjing conference as well as comments, S. Klose for all the financial support, and the contribution of the many authors who had a part in creating the two papers that this proceeding is based upon. Also thanks to Y. F. Huang for many an excellent information both before and after the conference.
\end{theacknowledgments}

% choose bibtex style depending on layout style and options used in
% sample:

\doingARLO[\bibliographystyle{aipproc}]
          {\ifthenelse{\equal{\AIPcitestyleselect}{num}}
             {\bibliographystyle{arlonum}}
             {\bibliographystyle{arlobib}}
          }
%\bibliography{Kann}

\end{document}